\RequirePackage{fixltx2e} 
\RequirePackage{xparse}
\documentclass[%
 reprint,
 amsmath,amssymb,
 aps,
pra,
]{revtex4-1}

\usepackage{graphicx}
\usepackage{dcolumn}
\usepackage{bm}

\usepackage{braket}
\usepackage{siunitx}
\usepackage{subcaption}
\usepackage{color}

\DeclareDocumentCommand\Diff{m}{\!\mathrm{d}#1\,}

\newcommand{\me}{\mathrm{e}}
\newcommand{\vac}{\ket{\text{vac}}}
\newcommand{\bvac}{\bra{\text{vac}}}
\newcommand{\kbvac}{\vac\!\bvac}

\DeclareDocumentCommand\aop{m}{\hat{a}_{#1}}

\DeclareDocumentCommand\aopd{m}{\hat{a}_{#1}^{\dagger}}

\newcommand{\comment}[1]{}

\begin{document}


\title{The effects of filtering on the purity of heralded single photons from parametric sources}

\author{Daniel R. Blay}
 \email{daniel.blay@mq.edu.au}
\author{M. J. Steel}%
\author{L. G. Helt}
\affiliation{%
 Macquarie University Quantum Research Centre in Science and Technology (QSciTech) and Centre for Ultrahigh bandwidth Devices for Optical Systems (CUDOS), MQ Photonics Research Centre, Department of Physics and Astronomy, Macquarie University, NSW 2109, Australia\\
}%

\date{\today}

\begin{abstract}
Optical quantum protocols are reliant on the production of highly pure single photons. However, most parametric heralded single photon sources possess spectral correlations, limiting the purity of the produced photon. The most common method for restoring the purity is with narrowband spectral filtering, which comes at the cost of reducing the rate of heralding detection. Here, we characterize this trade-off analytically for a large class of parametric sources. It is possible to achieve up to 20\% heralding success probabilities with 90\% heralded photon purities.
\end{abstract}

\pacs{Valid PACS appear here}
\maketitle


\section{Introduction}
\noindent Many optical quantum protocols would benefit from the use of highly pure single photons, produced on-demand at high rates. Presently there is no widespread source which fulfills this role, but the most mature platform for producing these photons is heralded single photon sources~\cite{Ursin2007,Scheel2009,Ma2012,Sangouard2012,Clark2015,Peiris2017}, which produce pairs of photons (``biphotons'') in nonlinear materials through parametric processes, including spontaneous parametric down-conversion (SPDC) and spontaneous four-wave mixing (SFWM). One of these photons is detected, ``heralding'' the presence of the other. These sources have many desirable properties, including high brightness and the ability to be integrated into existing optical technologies. Furthermore, we can draw on over fifty years of nonlinear optics research when designing these sources. This makes them promising sources for larger scale quantum information science experiments.

However, parametric sources of photon pairs also typically possess strong spectral correlations. Because this introduces classical uncertainty as to the spectral mode of the heralded single photon, the photon's purity is diminished. This limits the utility of such a source, as the quantum interference between independent heralded photons is consequently reduced. It is possible to engineer parametric sources to reduce these correlations~\cite{Grice2001,Mosley2008,Halder2009,Levine2010,Clark2011,Spring2013,Fang2013,Harder2013,Bruno2014,Weston2016}, however this can be challenging to implement. The most common and convenient solution is to take an existing source and add a narrowband filter to the heralding arm~\cite{Sharping2006,McMillan2009,Aboussouan2010,Sun2016,Valivarthi2016}, which suppresses the correlations at the expense of the rate at which photons are detected. Any means to optimize this compromise is desirable.

Previous studies have considered increasing single photon purities through filtering, in the spatial domain~\cite{VanExter2006} suited to bulk sources, as well as the spectral domain~\cite{Branczyk2010,Florez2015,Kamide2015,Du2015,Meyer-Scott2017a} more often considered for integrated sources. Similar techniques can be used to optimize two photon states for squeezing~\cite{Christ2014}. However, these studies have not fully explored the photon pair spectral correlation space, are not always well-connected to experiments, and tend to examine the problem through a single lens.

In this work, we build upon these previous efforts and calculate analytical expressions for several important quantities, focusing on the spectral regime most relevant to integrated single mode waveguide sources. In particular, we develop two methods for calculating the purity of the single photon resulting from detecting a filtered herald photon, as well as the probability of detecting said herald. Additionally, we calculate the shape of the Hong-Ou-Mandel (HOM) interference dip resulting from the interference of two such photons. In section~\ref{sec:fast_method}, we introduce our general formalism and present a method relying on direct integration over the biphoton state. This method lends itself well to finding analytical results and making direct connections between the aforementioned quantities and the physical parameters of the state. In section~\ref{sec:schmidt} we present a second method working in the spectral mode basis, utilizing the Schmidt decomposition~\cite{Law2004}. This second method illuminates the underlying multimode physics. Using these results, in section~\ref{sec:results} we explore the parameter space of a certain class of biphoton states and filters. We demonstrate that, even for intial states with low purities, favorable trade-offs between the purity and heralding success probability are achievable after filtering. We also suggest alternate schemes for increasing the purity of heralded photons that do not rely on filtering.

\section{Definitions and direct integration method}\label{sec:fast_method}
\subsection{Biphotons}
\begin{figure}[httb]
	\centering
	\includegraphics[width=0.75\linewidth]{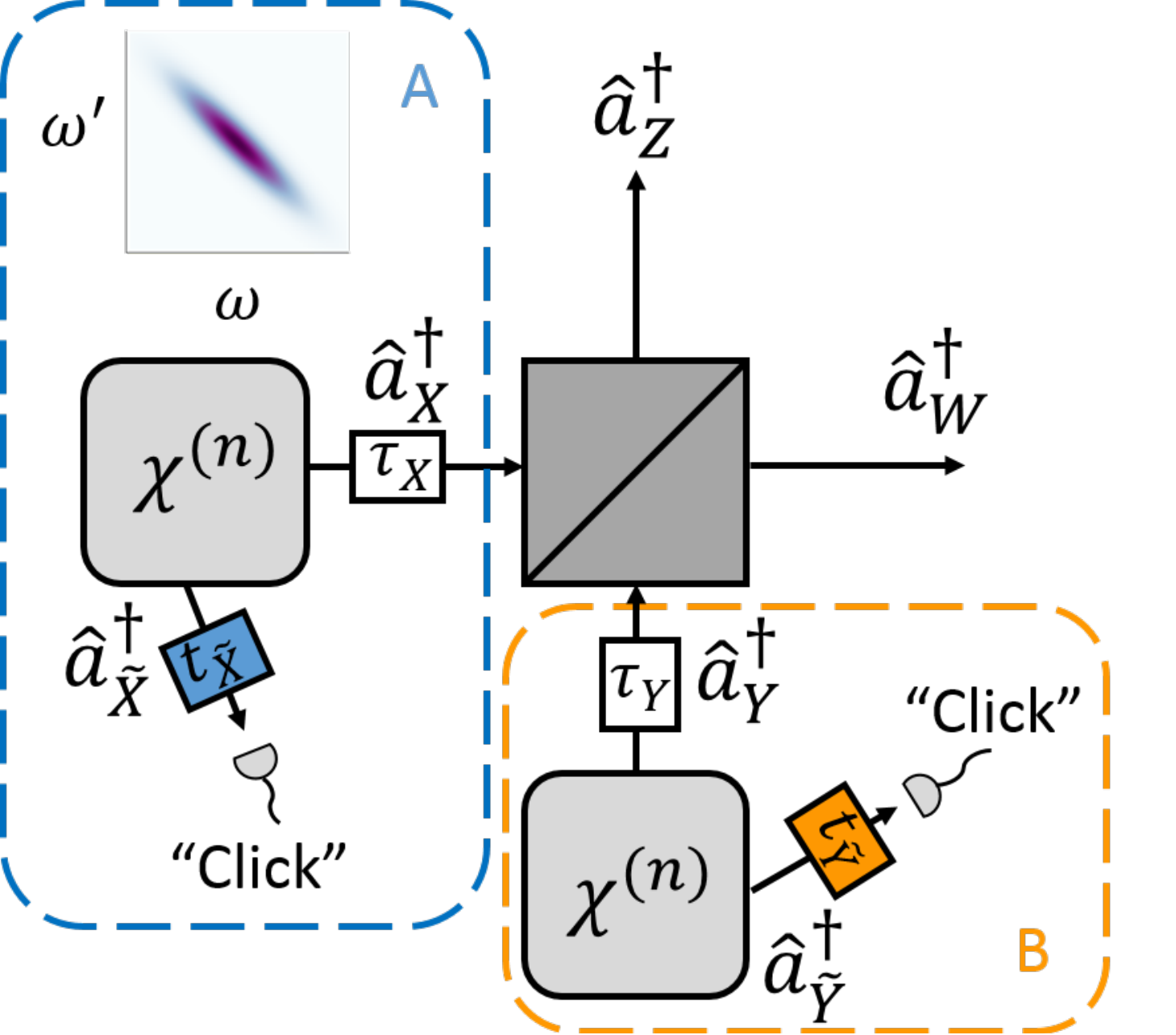}
	\caption{\label{fig:hom_exp} A diagrammatic representation of a two photon interference experiment. Here $\aopd{i}$ are creation operators corresponding to mode $i=(X,\widetilde{X},Y,\widetilde{Y},Z,W)$. The area within each dashed box indicates a single heralded photon source, labelled A or B, with a time delay of $\tau_i$ on the heralded arm. The shaded boxes on the heralded arms represent spectral filters. The inset image above the left source gives a typical example of the frequency correlations between signal and idler photons.}
\end{figure}

\noindent Parametric photon sources are spectrally multimode in nature, and as such any accurate description of their output states must also be multimode. These sources, such as the idealized ones shown in the dashed boxes in Fig. \ref{fig:hom_exp}, have output states comprised mostly of vacuum, with some higher order components including the two-photon component we are primarily concerned with. When the photons in the pair produced by the source can be discriminated in some degree of freedom (for example, polarization), the pair state can be expressed as
\begin{equation}\label{eq:pair_state}
\ket{\mathrm{II}}_A = \int \Diff{\omega}\,\Diff{\omega'} \Phi(\omega,\omega') \aopd{X}(\omega)\aopd{\widetilde{X}}(\omega')\me^{i \omega \tau_X}\vac.
\end{equation}
The phase term corresponds to a temporal delay $\tau_X$ between spatially distinct modes $X$ and $\widetilde{X}$, which becomes relevant when we consider HOM interference. The joint spectral amplitude (JSA) $\Phi(\omega,\omega')$ incorporates the multimode nature of the photon pair production. To allow for the consistent interpretation of the JSA as a probability amplitude we require ${\int \Diff{\omega}\,\Diff{\omega'} \left|\Phi(\omega,\omega')\right|^2 = 1}$. Often the JSAs associated with parametric processes such as SPDC and SFWM are highly correlated in frequency, as seen in the top of the left dashed box in Fig.~\ref{fig:hom_exp}. A heralded single photon generated from such a correlated state will have low purity, due to its projection into a mixture of frequency modes upon the detection of the herald photon~\cite{Branczyk2010}. Without any spectral filtering, this purity is given by~\cite{Fedorov2006}
\begin{align}
P =&\, \text{Tr}\left[\left(\ket{\mathrm{I}}_{A}\bra{\mathrm{I}}_{A}\right)^2\right]\nonumber\\
=&\, \int \Diff{\omega}\,\Diff{\omega'}\,\Diff{\omega''}\,\Diff{\omega'''} \Phi(\omega,\omega') \Phi^*(\omega'',\omega')\nonumber\\
&\times\Phi^*(\omega,\omega''') \Phi(\omega'',\omega'''),\label{eq:unfiltered_purity}
\end{align}
where
\begin{equation}
\ket{\text{I}}_A\bra{\text{I}}_A = \text{Tr}_{\widetilde{X}}\left[\ket{\mathrm{II}}_{A}\bra{\mathrm{II}}_{A}\right].
\end{equation}
When correlations are present, $P<1$. Spectral filtering can restore the purity.
\subsection{Heralding with filters}
\noindent In order to quantify the effects of filtering, we must first model a filter. We consider a filter as a frequency selective beamsplitter, with transmittance $t_{\widetilde{X}}(\omega)$ and reflectance $r_{\widetilde{X}}(\omega)$. Transmitted photons are those that make it to the detector, and the reflected photons are discarded. Then the measurement operator for the herald photon is given by
\begin{equation}\label{eq:measurement_operator}
\hat{F}_{\widetilde{X}} = \int \Diff{\omega} \left|t_{\widetilde{X}}(\omega)\right|^2 \aopd{\widetilde{X}}(\omega)\kbvac\aop{\widetilde{X}}(\omega),
\end{equation}
where the transmission is given by $\left|t_{\widetilde{X}}(\omega)\right|^2$. Conditioned on the photon in mode $\widetilde{X}$ being successfully detected, the reduced state is
\begin{align}
\hat{\rho}_X =&\, \frac{1}{\mathcal{S}}\text{Tr}_{\widetilde{X}}\left[\ket{\mathrm{II}}_A \bra{\mathrm{II}}_A \hat{F}_{\widetilde{X}}\right]\nonumber\\
=&\, \frac{1}{\mathcal{S}} \int \Diff{\omega}\,\Diff{\omega'}\,\Diff{\omega''} \left|t_{\widetilde{X}}(\omega')\right|^2 \Phi(\omega,\omega') \Phi^*(\omega'',\omega')\nonumber\\
&\cdot\aopd{X}(\omega)\kbvac\aop{X}(\omega'') \me^{i(\omega-\omega'')\tau_X}\label{eq:heralded_single_photon_state}
\end{align}
The normalization constant $\mathcal{S}$ corresponds to the probability of a successful herald detection, which is found by the requirement that $\text{Tr}\left[\hat{\rho}_{X}\right] = 1$, giving
\begin{equation}\label{eq:detection_probability}
\mathcal{S} = \int \Diff{\omega}\,\Diff{\omega'} \left|t_{\widetilde{X}}(\omega')\right|^2 \left|\Phi(\omega,\omega')\right|^2.
\end{equation}
The heralded single photon purity is then simply
\begin{align}
\mathcal{P} =&\, \text{Tr}\left[\hat{\rho}_X^2\right]\nonumber\\
=&\, \frac{1}{\mathcal{S}^2} \int \Diff{\omega}\,\Diff{\omega'}\,\Diff{\omega''}\,\Diff{\omega'''} \left|t_{\widetilde{X}}(\omega')\right|^2 \left|t_{\widetilde{X}}(\omega''')\right|^2\nonumber\\
&\times\Phi(\omega,\omega') \Phi^*(\omega'',\omega') \Phi^*(\omega,\omega''') \Phi(\omega'',\omega''') \label{eq:filtered_purity}
\end{align}
In the limit where the filter has perfect transmission over the spectral band of the photon (e.g. $\left|t_{\widetilde{X}}(\omega)\right|^2=1$), this reduces simply to the unfiltered purity found in Eq.~(\ref{eq:unfiltered_purity}), $\mathcal{P} = P$, and $\mathcal{S} = 1$. For a sufficiently narrowband filter, $\mathcal{P} \to 1$. However, in this limit, $\mathcal{S} \to 0$ (in section \ref{sec:schmidt} we provide a Schmidt-mode perspective on these limits). In section \ref{sec:results}, we address the question of whether there exists a regime between these limits which allows for the useful operation of the source.

\subsection{Hong-Ou-Mandel interference}
\noindent As purities are often characterized in two-photon interference experiments, it is useful to consider a HOM set-up as shown in Fig.~\ref{fig:hom_exp}. Two identical sources $A$ and $B$ generate biphotons $\ket{\mathrm{II}}_A$ and $\ket{\mathrm{II}}_B$, such that the initial state is
\begin{equation}
\hat{\rho}_\text{in} = \ket{\mathrm{II}}_A\bra{\mathrm{II}}_A \otimes \ket{\mathrm{II}}_B\bra{\mathrm{II}}_B.
\end{equation}
Detecting the herald photons corresponding to $\aopd{\widetilde{X}}(\omega)$ and $\aopd{\widetilde{Y}}(\omega)$ leaves the system in the space spanned by mode operators $\aopd{X}(\omega)$ and $\aopd{Y}(\omega)$, appropriate for HOM interference. If there are strong spectral correlations between the herald and heralded photons, the resulting interference visibility will be poor due to the low heralded photon purity. For a beamsplitter with reflection $R$ and transmission $T$ (independent of frequency), this visibility is given simply by ${V = RTP/[1-(2+P)RT]}$~\cite{Loudon2000}, where $P$ is the unfiltered purity found in Eq.~(\ref{eq:unfiltered_purity}). How does this expression change when we consider filtering?

When both photons are successfully heralded when filtered, the resulting reduced state is
\begin{align}
\hat{\rho}_{XY} =&\, \mathcal{S}_{\widehat{X}}^{-1}\mathcal{S}_{\widehat{Y}}^{-1}\, \text{Tr}_{\widetilde{X} \widetilde{Y}}\left[\hat{\rho}_\text{in} \hat{F}_{\widetilde{X}} \hat{F}_{\widetilde{Y}}\right]\nonumber\\
=&\, \mathcal{S}_{\widehat{X}}^{-1}\mathcal{S}_{\widehat{X}}^{-1} \int \Diff{\omega_{\widetilde{X}}}\,\Diff{\omega_{\widetilde{Y}}}\,\Diff{\omega_1}\,\Diff{\omega_2}\,\Diff{\omega_1'}\,\Diff{\omega_2'}\nonumber\\
&\times\left|t_{\widetilde{X}}(\omega_{\widetilde{X}})\right|^2 \Phi(\omega_1,\omega_{\widetilde{X}}) \Phi^*(\omega_1',\omega_{\widetilde{X}})\nonumber\\
&\times\left|t_{\widetilde{Y}}(\omega_{\widetilde{Y}})\right|^2\Phi^*(\omega_2',\omega_{\widetilde{Y}}) \Phi(\omega_2,\omega_{\widetilde{Y}})\nonumber\\
&\cdot \aopd{X}(\omega_1) \aopd{Y}(\omega_2) \kbvac \aop{X}(\omega_1')\aop{Y}(\omega_2')\nonumber\\ &\times\me^{i\tau_X(\omega_1-\omega_1') + i\tau_Y(\omega_2-\omega_2')},\label{eq:heralded_two_photon_state}
\end{align}
where $\mathcal{S}_i$ is given by Eq.~(\ref{eq:detection_probability}) for the appropriate filter function $t_i(\omega)$. As we are interested in HOM interference dips, we perform the beamsplitter transformation,
\begin{align}
\aopd{X}(\omega) &\to t \aopd{W}(\omega) + r \aopd{Z}(\omega)\\
\aopd{Y}(\omega) &\to r' \aopd{W}(\omega) + t \aopd{Z}(\omega)
\end{align}
where $r$ and $r'$ are reflectances and $t$ the transmittance. We note that while $\left|r\right| = \left|r'\right|$, they have different phases. The output state is transformed as $\hat{\rho}_{XY} \to \hat{\rho}_{WZ}$. The probability of detecting a photon at both output arms (a coincidence) is given by
\begin{align}\label{eq:coincidences}
\braket{\hat{n}_Z\hat{n}_W} =&\, \text{Tr}\left[\hat{\rho}_{ZW}\hat{n}_Z\hat{n}_W\right]\nonumber\\
=&\, 1 - 2R T \Big(
1 + \mathcal{S}^{-2} \int \Diff{\omega_{\widetilde{X}}}\,\Diff{\omega_{\widetilde{Y}}}\,\Diff{\omega_W}\,\Diff{\omega_Z}\nonumber\\
&\times \left|t_{\widetilde{X}}(\omega_{\widetilde{X}})\right|^2 \left|t_{\widetilde{Y}}(\omega_{\widetilde{Y}})\right|^2 \Phi(\omega_W,\omega_{\widetilde{X}}) \Phi^*(\omega_Z,\omega_{\widetilde{X}})\nonumber\\
&\times \Phi^*(\omega_W,\omega_{\widetilde{Y}}) \Phi(\omega_Z,\omega_{\widetilde{Y}}) \me^{i(\omega_W-\omega_Z)(\tau_X-\tau_Y)}
\Big),
\end{align}
where $R = \left|r\right|^2$ and $T = \left|t\right|^2$ are the reflection and transmission coefficients, and the total number operator is
\begin{equation}
\hat{n}_i = \int\Diff{\omega}\aopd{i}(\omega)\aop{i}(\omega)
\end{equation}
Equation~(\ref{eq:coincidences}) describes the shape of the HOM interference dip as a function of temporal delay $\Delta\tau=\tau_X - \tau_Y$. In the appropriate limits (for identical filters $t_{\widetilde{X}}(\omega)=t_{\widetilde{Y}}(\omega)$), this equation becomes
\begin{align}
\braket{\hat{n}_Z\hat{n}_W}_{\Delta\tau \to 0} =&\, 1 - 2 R T\left(1+\mathcal{P}\right),\\
\braket{\hat{n}_Z\hat{n}_W}_{\Delta\tau \to \infty} =&\, 1 - 2 R T,
\end{align}
and so we can express the visibility of the HOM interference fringe as
\begin{align}\label{eq:visibility}
V &\equiv \frac{\braket{\hat{n}_Z\hat{n}_W}_{\Delta\tau \to \infty}-\braket{\hat{n}_Z\hat{n}_W}_{\Delta\tau \to 0}}{\braket{\hat{n}_Z\hat{n}_W}_{\Delta\tau \to \infty}+\braket{\hat{n}_Z\hat{n}_W}_{\Delta\tau \to 0}}\nonumber\\
&= \frac{RT \mathcal{P}}{1-\left(2+\mathcal{P}\right)RT}.
\end{align}
This has the same form as the unfiltered case, only with the heralded single photon purity accounting for the filtering, and $\mathcal{P}\geq P$.

\subsection{Analytical expressions}\label{sec:analytics}
\noindent In this section, we develop closed form expressions for the purity $\mathcal{P}$, the heralding success probability $\mathcal{S}$ and the two-photon interference visibility $V$.

\begin{figure}[httb]
	\centering
	\includegraphics[width=0.5\linewidth]{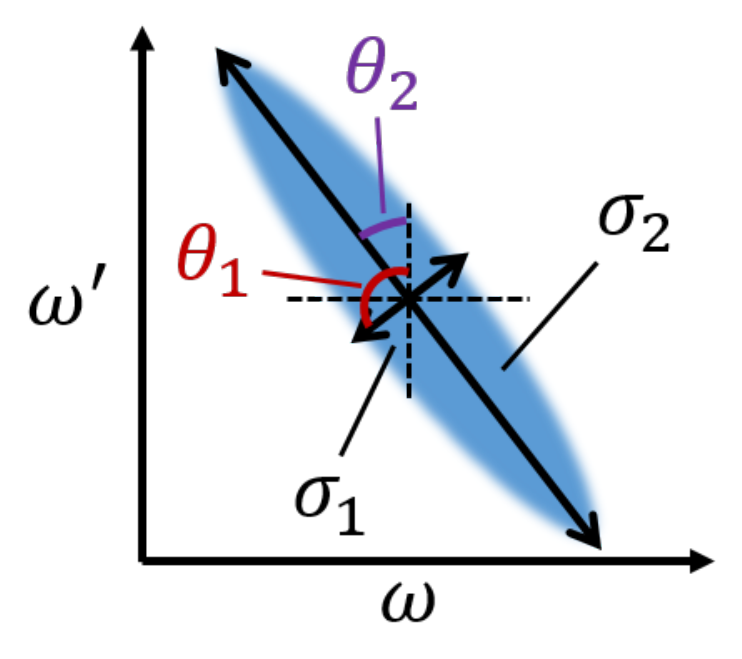}
	\caption{A cartoon of a double-Gaussian JSA, with the parameters $(\sigma_1,\sigma_2,\theta_1,\theta_2)$ labeled. Here $\sigma_1$ and $\sigma_2$ denote the widths of their respective Gaussians, oriented at $\theta_1$ and $\theta_2$.\label{fig:diagram}}
\end{figure}

Most JSAs have the form ${\Phi(\omega,\omega')\propto \phi(\omega+\omega')\text{sinc}\left(\Delta k L/2\right)}$, where $\phi(\omega)$ is the pump's spectral envelope and the sinc function accounts for phasematching. We approximate JSAs of this form as double-Gaussians,
\begin{multline}\label{eq:double_gaussian}
\Phi(\omega,\omega') =\\ \sqrt{\frac{\left|\sin(\theta_1-\theta_2)\right|}{\pi \sigma_1 \sigma_2}} \exp\left[-\left(\frac{\omega \sin\theta_1 + \omega' \cos\theta_1}{\sqrt{2}\sigma_1}\right)^2\right]\\
\times\exp\left[-\left(\frac{\omega \sin\theta_2 + \omega' \cos\theta_2}{\sqrt{2}\sigma_2}\right)^2\right].
\end{multline}
In this approximation, $\sigma_1$ and $\sigma_2$ are the widths of their respective Gaussians, and $\theta_1$ and $\theta_2$ are their orientations, where $\theta_1 \neq \theta_2$. A diagrammatic representation of these parameters can be found in Fig.~\ref{fig:diagram}. Although this approximation does not account for the lobes generated by the sinc phasematching term, it is nonetheless a fair approximation for many JSAs~\cite{Fedorov2006,Christ2011,Christ2014}.

If the filter is Gaussian,
\begin{equation}
\left|t_{\widetilde{X}}(\omega)\right|^2 = \exp\left[-(\omega-\omega_0)^2/(2\sigma_f^2)\right]\label{eq:gaussian_filter},
\end{equation}
where $\omega_0$ is the center frequency of the transmission filter, and $\sigma_f$ is the filter bandwidth, then we can find closed form expressions for the purity and heralded success probability. We have
\begin{multline}
\mathcal{S} = \sqrt{2\sigma_f^2 \epsilon^{-1} \sin^2\left(\theta_1-\theta_2\right)}\exp\left[-\frac{\omega_0^2\sin^2\left(\theta_1-\theta_2\right)}{\epsilon}\right]
\end{multline}
and
\begin{equation}\label{eq:analytical_filtered_purity}
\mathcal{P} = \sqrt{\frac{{\sigma_1^2\sigma_2^2\zeta}}{2\left(2\sigma_2^2\sigma_f2\cos^2\theta_1 + \sigma_1^2\xi\right)\left(\sigma_1^2\sin^2\theta_2+\sigma_2^2\sin^2\theta_1\right)}}
\end{equation}
where
\begin{align}
\epsilon =&\, \sigma_1^2\sin^2\theta_2 + \sigma_2^2\sin^2\theta_1 + 2\sigma_f^2 \sin^2\left(\theta_1-\theta_2\right),\\
\zeta =&\, \sigma_1^2 + \sigma_2^2  2\sigma_f^2 - \sigma_1^2\cos\left(2\theta_2\right) - \sigma_2^2\cos\left(2\theta_1\right)\nonumber\\
&- 2\sigma_f^2\cos\left[2\left(\theta_1-\theta_2\right)\right]
\end{align}
and
\begin{equation}
\xi = \sigma_2^2 + 2\sigma_f^2 \cos^2\theta_2.
\end{equation}
Finally, when the filters on both sources are identical, we can express the HOM dip analytically as
\begin{multline}\label{eq:hom_shape_an}
\braket{\hat{n}_W \hat{n}_Z} =\\ 1 - 2 RT\left(1 + \mathcal{P} \exp\left[-\frac{\Delta\tau^2 \sigma_1^2 \sigma_2^2}{2\left(\sigma_1^2\sin^2\theta_2 + \sigma_2^2\sin^2\theta_1\right)}\right]\right).
\end{multline}
We note that although the visibility as given by Eq.~(\ref{eq:visibility}) depends on the filter width through $\mathcal{P}$, the shape of the HOM dip as shown in Eq.~(\ref{eq:hom_shape_an}) is otherwise independent of the filter width. This says that while the depth of the HOM dip depends on the filter width, the width (e.g. full-width at half-depth) of the HOM dip does not. This occurs as the heralding spectral filter does not change the characteristic width of the JSA along the $\omega$ (rather than $\omega'$, see Fig. \ref{fig:diagram}) axis, which the heralded photon inherits. In the case where a spectral filter is present on the heralded arm, the inherited spectrum of the heralded photon does change, and thus so do its temporal properties (e.g. a narrower spectral filter yields a wider HOM dip). The expressions for the purity in Eq.~(\ref{eq:filtered_purity}) and the heralding success probability in Eq.~(\ref{eq:detection_probability}) have a closed form in terms of the parameters of the double-Gaussian and the filter.

This method has provided us with analytical results of great utility. However, the details of the underlying physics are not clear with the integral approach.

\section{A Schmidt mode perspective}\label{sec:schmidt}
\noindent In order to tease out the multimode physics inherent to parametric photon sources, we now take an alternate approach to analyze the heralded single photon purity $\mathcal{P}$ and detection probability $\mathcal{S}$. This method requires that we take the Schmidt decomposition of the JSA. This decomposition is expressed as
\begin{equation}\label{eq:schmidt_decomp}
\Phi(\omega,\omega') = \sum_{\mu} p_\mu \Gamma_{\mu}(\omega) \Theta_{\mu}(\omega'),
\end{equation}
where the Schmidt functions $\Gamma_{\mu}$ and $\Theta_{\mu}$ correspond to the spectral modes that constitute the JSA, and form a complete and orthonormal basis (see appendix for details). The Schmidt coefficients are strictly positive and must satisfy
\begin{equation}
\sum_{\mu} p_\mu = 1,
\end{equation}
and furthermore the unfiltered purity (recalling Eq.~(\ref{eq:unfiltered_purity})) is related simply to the Schmidt number $K$ through
\begin{equation}\label{eq:schmidt_number}
P = 1/K \equiv \sum_{\mu} p_\mu^2.
\end{equation}
Equations~(\ref{eq:schmidt_decomp}) and (\ref{eq:schmidt_number}) make it clear that when the JSA is separable (i.e. can be expressed as a product of two functions), the Schmidt decomposition contains a single term ($p_\mu = \delta_{\mu,1}$) and so $P = K = 1$. In all other cases, $P < 1$ and $K>1$. Note that when the probability of pair production is small, the Schmidt number is also connected with the time-independent second-order correlation function~\cite{Christ2011} $g^{(2)} = 1 + 1/K$. When the JSA is uncorrelated (separable), we should expect $g^{(2)} = 2$.

By direct substitution into Eqs. (\ref{eq:filtered_purity}) and (\ref{eq:detection_probability}) the purity and heralding detection probability are given by
\begin{align}
\mathcal{P} &= \frac{1}{\mathcal{S}^2} \sum_{\mu\nu} \left|Q_{\mu\nu}\right|^2 p_\mu p_\nu,\\
\mathcal{S} &= \sum_{\mu} Q_{\mu\mu} p_\mu,
\end{align}
where the overlaps with the filter and Schmidt functions,
\begin{equation}
Q_{\mu\nu} = \int \Diff{\omega} \left|t_{\widetilde{X}}(\omega)\right|^2 \Theta_{\mu}(\omega) \Theta_{\nu}^*(\omega),
\end{equation}
provide a measure of the degree of mode-mixing introduced by the filter -- the filter removes the orthonormality of the Schmidt modes. In this formulation, it is easy to see that when the filter has perfect transmission over the band of the photon (e.g. $\left|t_{\widetilde{X}}(\omega)\right|^2=1$), the overlaps reduce to $Q_{\mu\nu} = \delta_{\mu,\nu}$, and thus the purity reduces to the unfiltered form $\mathcal{P} = \sum_{\mu} p_\mu^2$, and $\mathcal{S} = 1$. For a sufficiently narrowband filter $\mathcal{P}\to1$. However, in this limit, $\mathcal{S}\to0$, making the use of such a filter impractical.

With the Schmidt decomposition the shape of the HOM dip is
\begin{align}
\braket{\hat{n}_Z \hat{n}_W} =& 1 - 2 RT\bigg(1 + \mathcal{S}^{-2} \sum_{\mu \nu \mu' \nu'} Q_{\mu \mu'} Q_{\nu \nu'}\nonumber\\ &\times\sqrt{p_\mu p_\nu p_{\mu'} p_{\nu'}} \int \Diff{\omega_W}\Diff{\omega_Z} \Gamma^*_{\mu'}(\omega_W) \Gamma_{\nu}(\omega_W)\nonumber\\
&\times\Gamma^*_{\nu'}(\omega_Z) \Gamma_{\mu}(\omega_Z) \me^{i(\omega_W-\omega_Z)(\tau_X - \tau_Y)}
\bigg).\label{eq:schmidt_hom_dip}
\end{align}
This makes it eminently clear that the shape of the HOM dip cannot depend on the width of the filter, which only appears in the overlaps $Q_{\mu\nu}$, whereas the width of the HOM dip is determined by the integrals in Eq.~(\ref{eq:schmidt_hom_dip}). Another advantage of these expressions is how easily they are evaluated numerically. Whilst the Schmidt coefficients and functions can be found analytically for double-Gaussian JSAs (see appendix), we note that they can always be found numerically from a singular value decomposition (SVD) of a discretized JSA.

\section{Results and Discussion}\label{sec:results}
\subsection{Filtering double-Gaussians}
\begin{figure}[httb]
	\centering
	\begin{subfigure}{.9\linewidth}
		\centering
		\includegraphics[width=\linewidth]{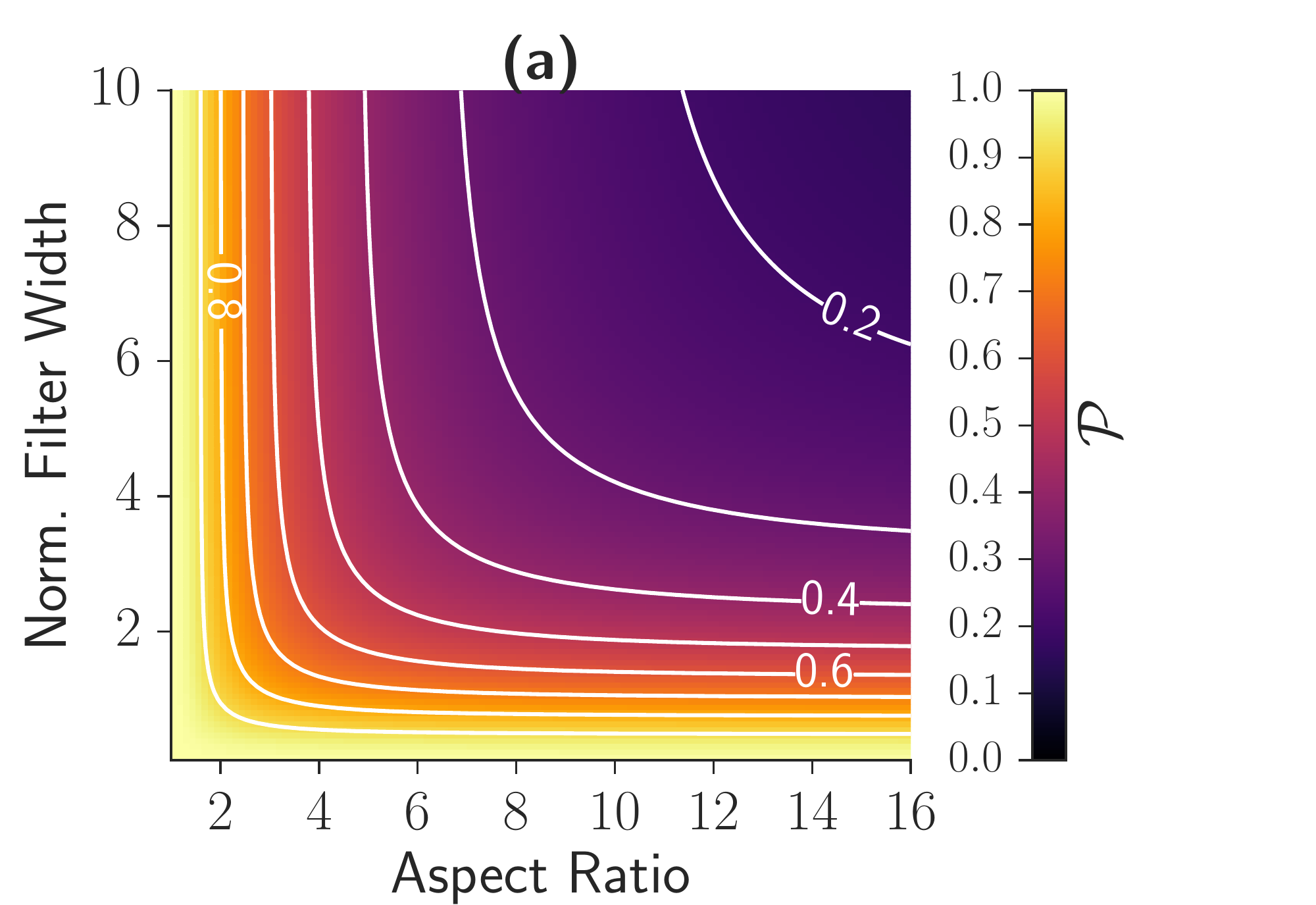}
	\end{subfigure}
	\begin{subfigure}{.9\linewidth}
		\centering
		\includegraphics[width=\linewidth]{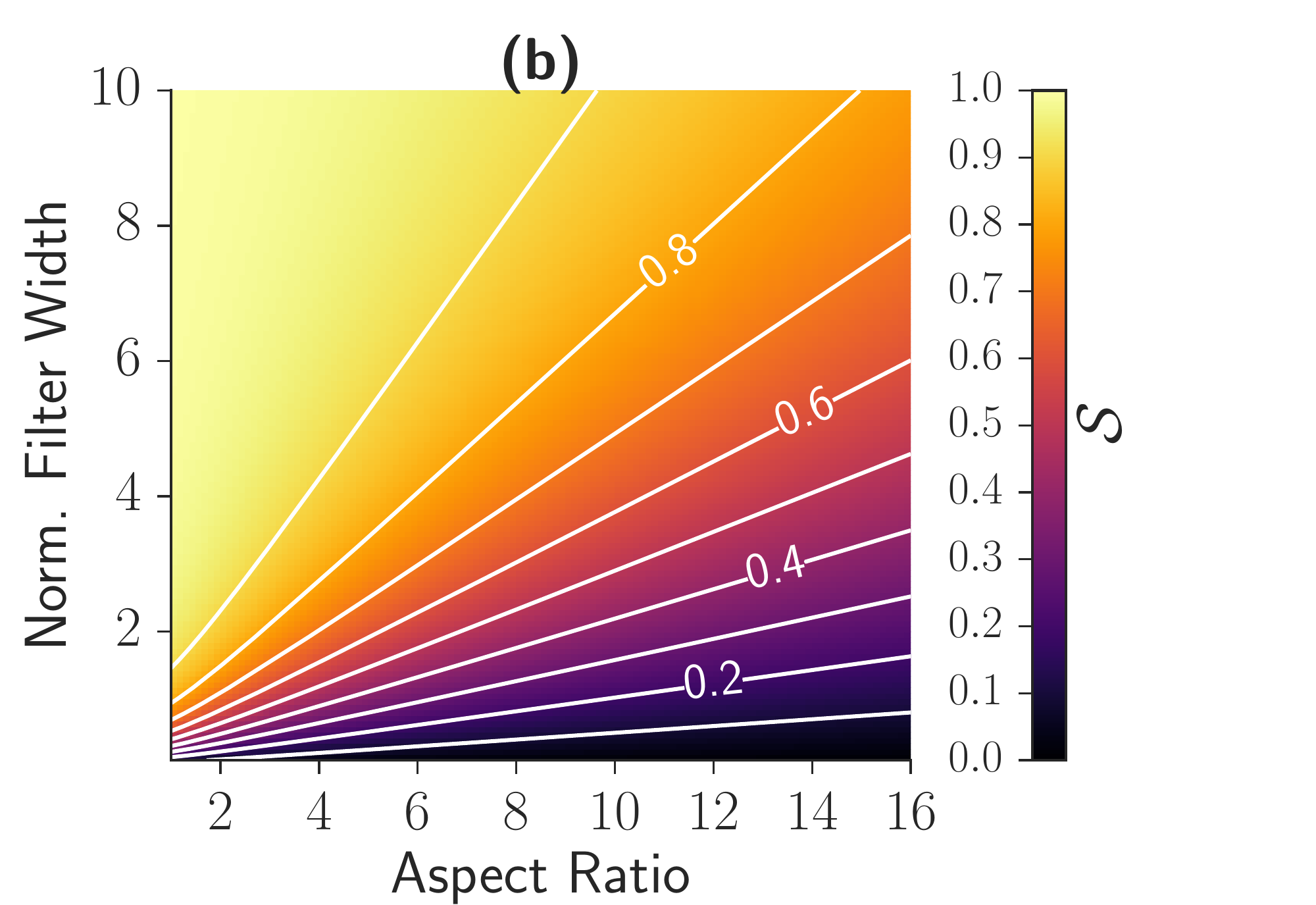}
	\end{subfigure}
	\caption{\textbf{(a)}: The heralded single photon purity $\mathcal{P}$ for a JSA with a varying aspect ratio ($\sigma_2/\sigma_1$), and normalized filter widths ($\sigma_f/\sigma_1$). \textbf{(b)}: The heralding success probability $S$ for a JSA with a varying aspect ratio, and normalized filter widths. In this figure, ${\theta_1 = \pi/4}$, ${\theta_2 = -\pi/4}$ and $\sigma_1=1$.\label{fig:aspect_ratio_plots}}
\end{figure}

\begin{figure}[httb]
	\centering
	\includegraphics[width=0.8\linewidth]{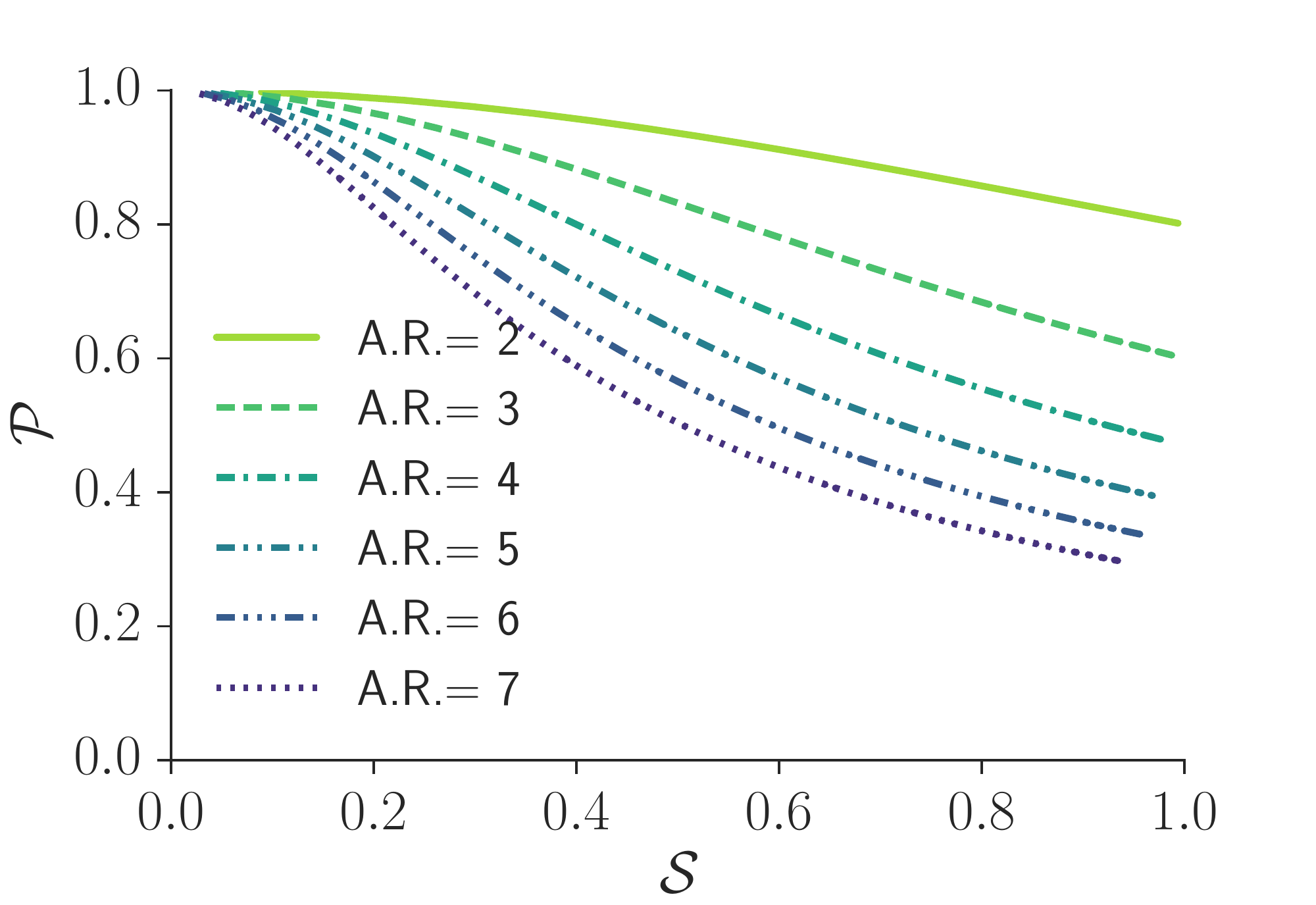}
	\caption{The trade-off between heralded single photon purity and efficiency for several aspect ratios ($\text{A.R.}=\sigma_2/\sigma_1$). We note that up to $\sigma_f/\sigma_1\approx6$, high purities $\mathcal{P}>0.9$ can be achieved for $\mathcal{S}\sim0.2$.\label{fig:slice_S_P}}
\end{figure}

\begin{figure}[httb]
	\centering
	\begin{subfigure}{.9\linewidth}
		\centering
		\includegraphics[width=\linewidth]{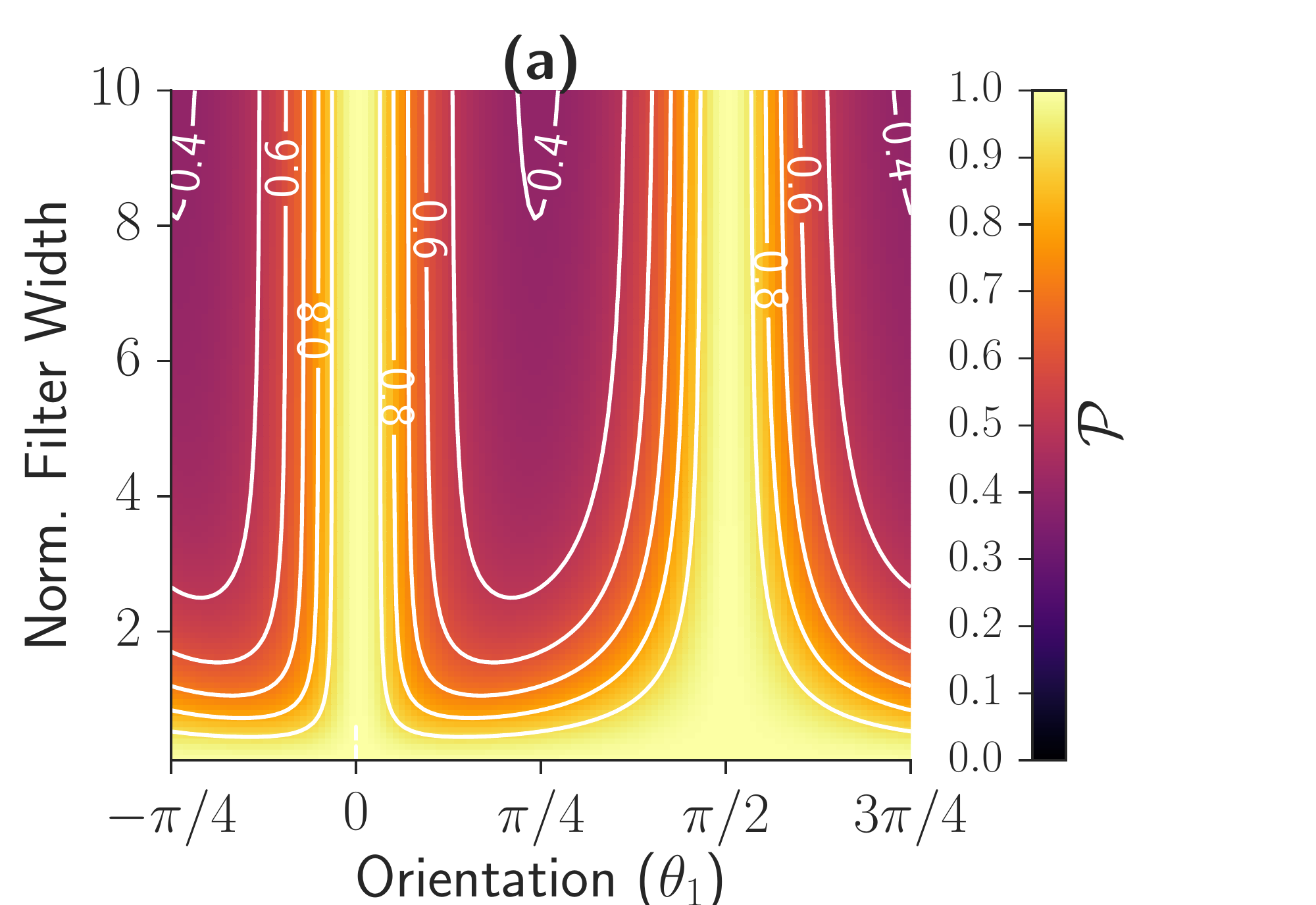}
	\end{subfigure}
	\begin{subfigure}{.9\linewidth}
		\centering
		\includegraphics[width=\linewidth]{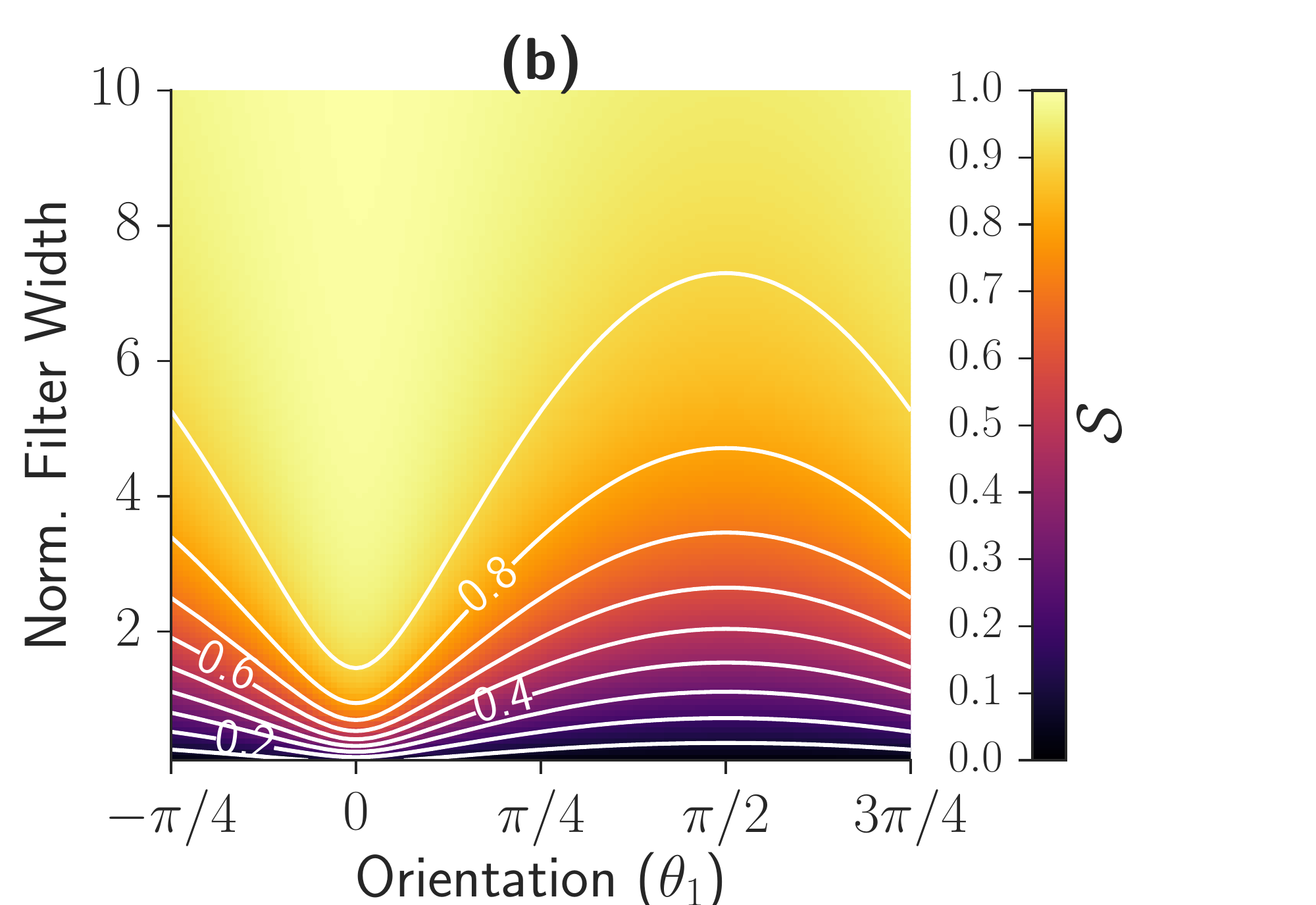}
	\end{subfigure}
	\caption{\textbf{(a)}: The heralded single photon purity $\mathcal{P}$ for a JSA with varying orientation ($\theta_1$ with ${\theta_2=\theta_1-\pi/2}$), and normalized filter widths ($\sigma_f/\sigma_1$). \textbf{(b)}: The heralding success probability $\mathcal{S}$ for a JSA with varying orientation, and normalized filter widths. In this figure, $\sigma_2=5$ and $\sigma_1=1$.\label{fig:orientation_plots}}
\end{figure}

\begin{figure}[httb]
	\centering
	\includegraphics[width=0.8\linewidth]{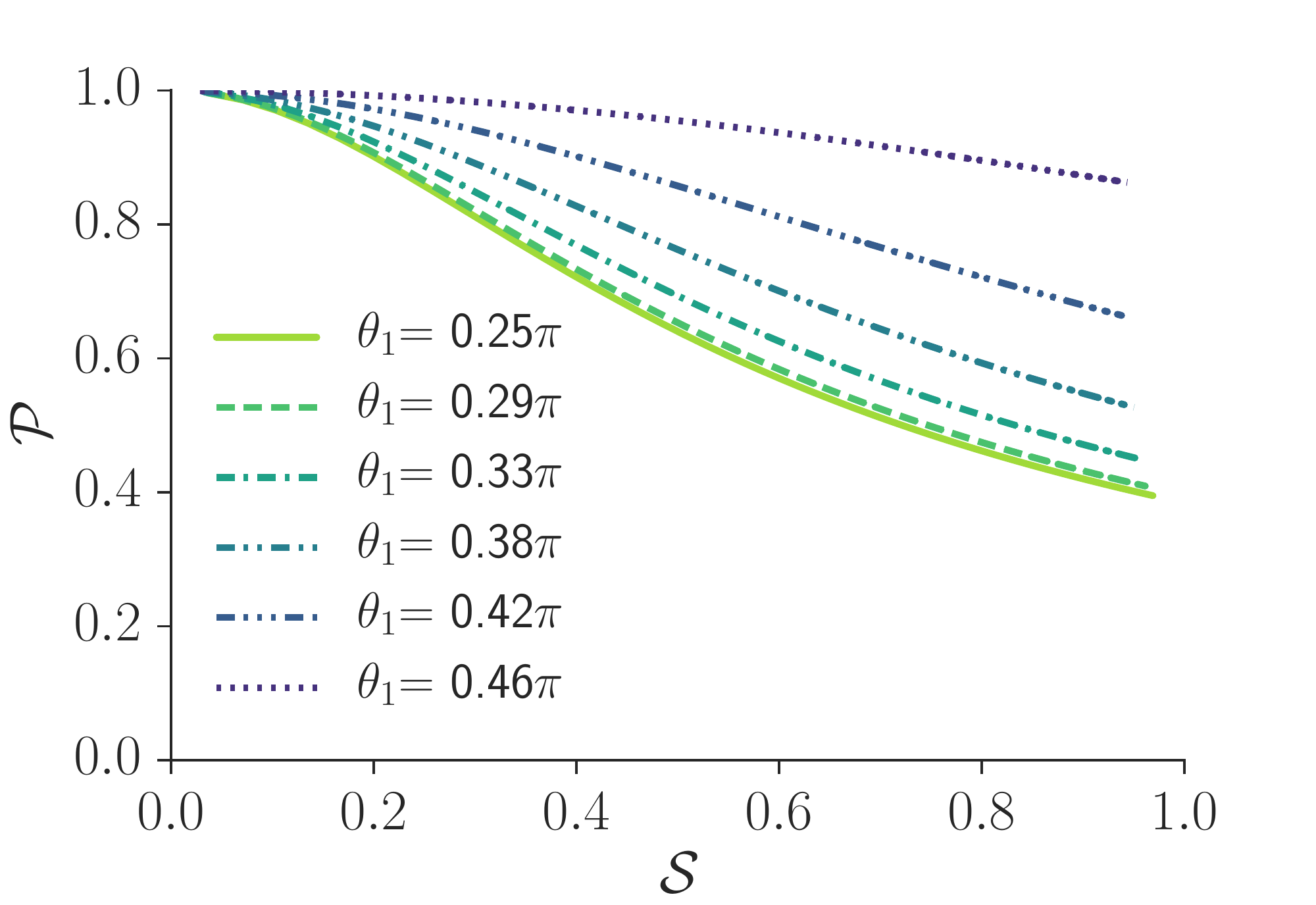}
	\caption{The trade-off between heralded single photon purity and efficiency for several orientations. We note that for orientations between $0.25\pi$ and $0.42\pi$, high purities $\mathcal{P}>0.9$ can be achieved for $\mathcal{S}\sim0.2$.\label{fig:Orientation_slice_S_P}}
\end{figure}

\noindent Using these analytical results, we now explore the parameter space of double-Gaussian JSAs [recall Eq.~(\ref{eq:double_gaussian})] $(\sigma_1,\sigma_2,\theta_1,\theta_2)$ and Gaussian filter widths $\sigma_f$. In Fig.~\ref{fig:aspect_ratio_plots}, we consider the trade-off between the heralded single photon purity $\mathcal{P}$ and the probability of successful herald detection $\mathcal{S}$ for a set orientation (${\theta_1 = \pi/4,\,\theta_2=-\pi/4}$), with varying aspect ratios ($\sigma_2/\sigma_1$) and normalized filter widths ($\sigma_f/\sigma_1$). As expected from Eqs.~(\ref{eq:double_gaussian}) and (\ref{eq:analytical_filtered_purity}), $\mathcal{P}=1$ for any filter if $\sigma_2/\sigma_1=1$ as there is only a single Schmidt mode, and $\mathcal{P}\to1$ as $\sigma_f/\sigma_1\to0$. Something that is not clear directly from Eqs.~(\ref{eq:filtered_purity}) and (\ref{eq:detection_probability}) is that high purities $\mathcal{P}>0.9$ can be achieved for successful detection probabilities of $\mathcal{S}\sim0.2$, even for aspect ratios as high as 6. This is illustrated in Fig.~\ref{fig:slice_S_P}.
Similarly, we can choose an aspect ratio (for Fig.~\ref{fig:orientation_plots}, ${\sigma_2/\sigma_1=5}$), set $\theta_2 = \theta_1 - \pi/2$ and vary the orientation of the JSA. Generally we do not require ${\theta_1+\theta_2=\pi/2}$, however this restriction makes the parameter space easier to examine. In the case where $\theta_1 = 0, \pi/2$ (\emph{i.e.} the JSA is oriented vertically or horizontally), we find that $\mathcal{P}=1$ for any filter width, as the Schmidt decomposition only has a single term. Outside this regime, we again discover that heralded single photons of high purity $\mathcal{P}>0.9$ are available for $\mathcal{S}\sim0.2$, as seen in Fig.~\ref{fig:Orientation_slice_S_P}. We note that the high $\mathcal{S}$ in Fig.~\ref{fig:orientation_plots}.b near $\theta_1=0$ is the result of the long axis of the JSA being perpendicular to the Gaussian filter. As a result, not as much light is filtered out as compared with JSAs of other orientations, leading to a higher heralding success probability.

\begin{figure}[httb]
	\centering
	\includegraphics[width=0.8\linewidth]{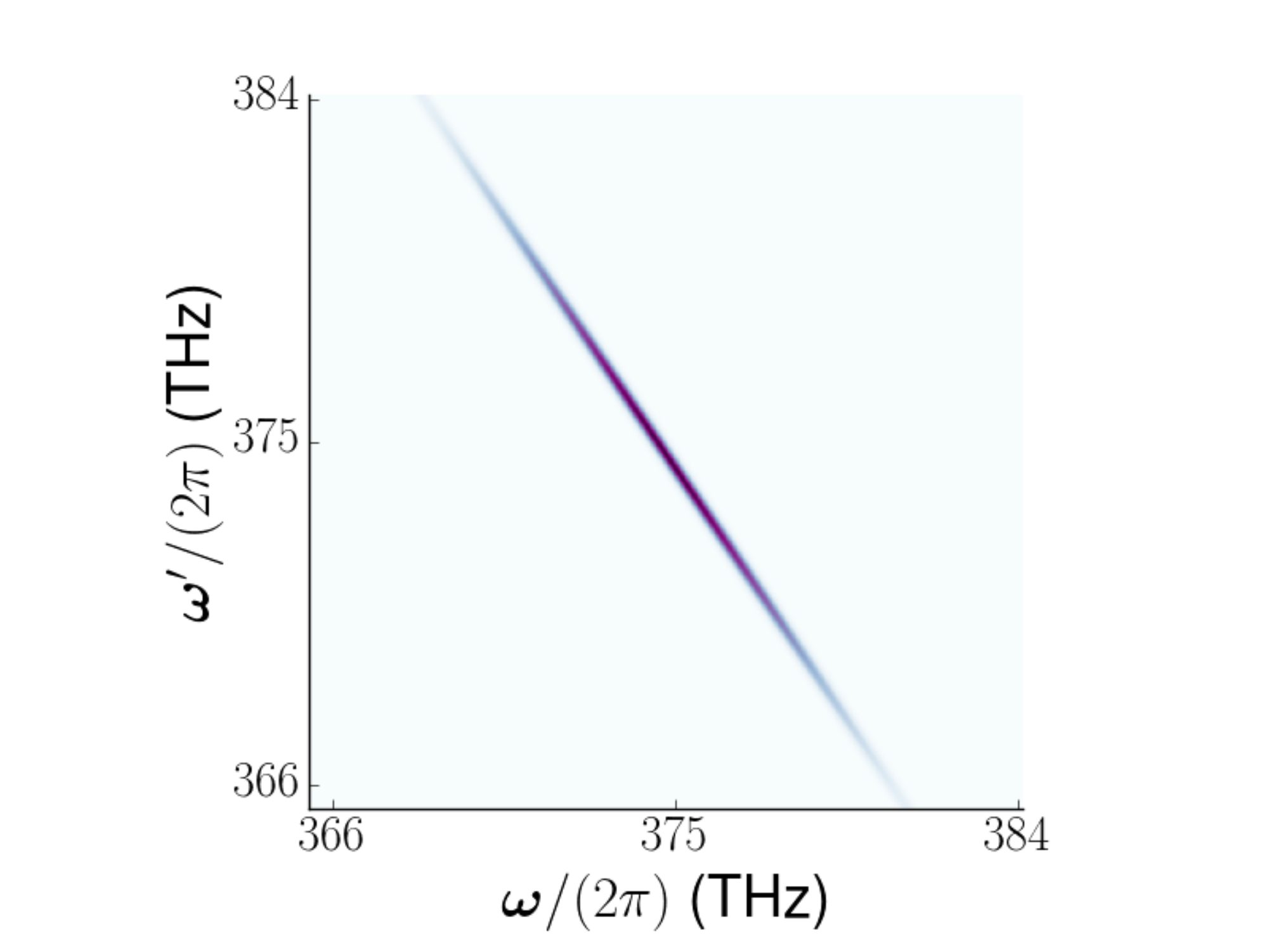}
	\caption{A double-Gaussian JSA with parameters ${(\sigma_1,\sigma_2,\theta_1,\theta_2) = (\SI{6.0}{\per\pico\second},\SI{0.70}{\per\pico\second},\SI{\pi/4}{\radian},\SI{0.97}{\radian})}$, corresponding to the source in \cite{Branczyk2010}.\label{fig:fitted_jsa}}
\end{figure}

\begin{figure}[httb]
	\centering
	\includegraphics[width=0.8\linewidth]{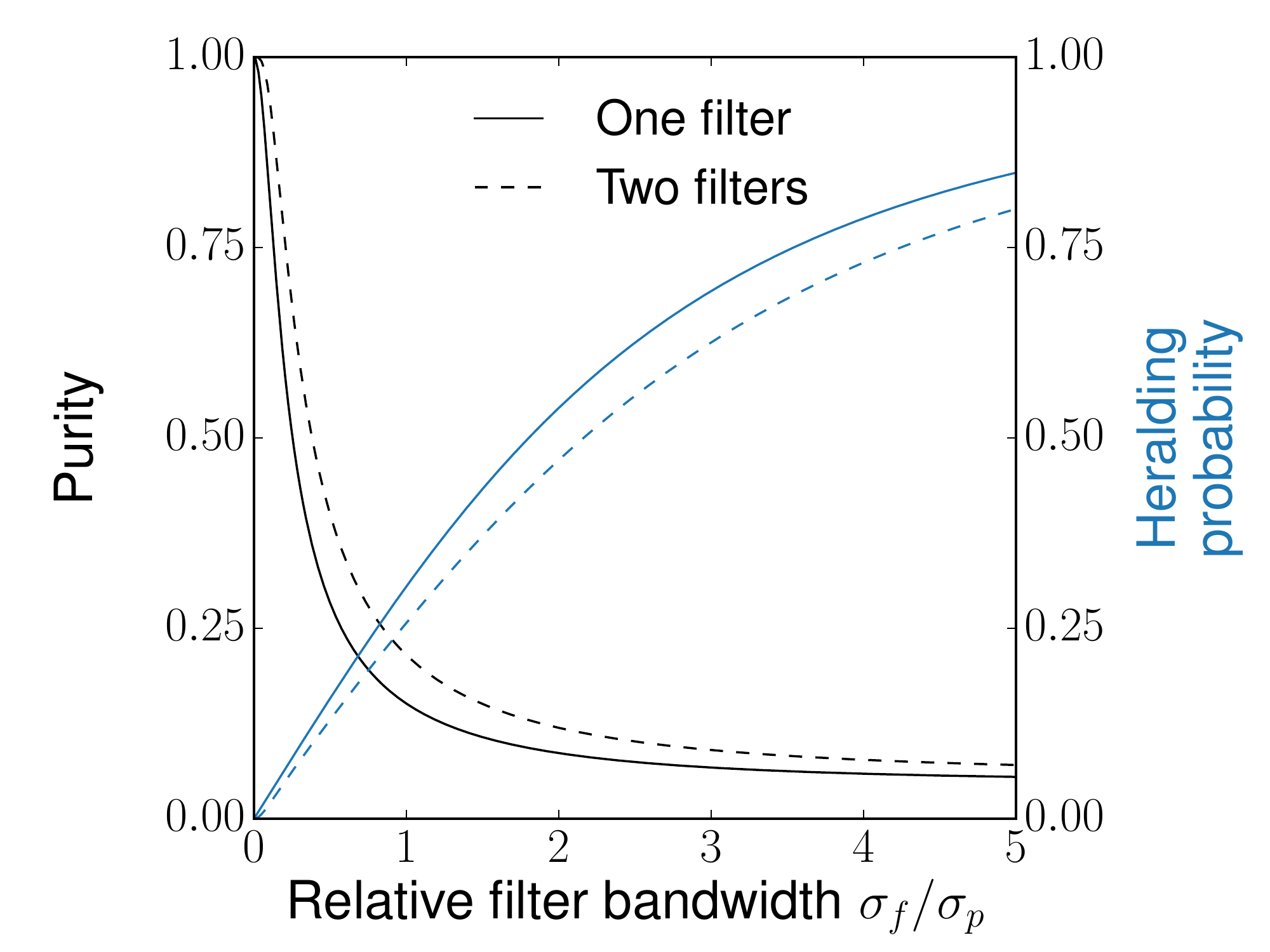}
	\caption{The heralded photon purity (black) and heralding success probability (blue/gray) for the JSA as shown in Fig.~\ref{fig:fitted_jsa}. The solid lines show $\mathcal{P}$ and $\mathcal{S}$ for a single filter, and the dashed lines for two filters ($\mathcal{P}_2$ and $\mathcal{S}_2$). The bandwidth of the filter being matched to the pump bandwidth occurs for $\sigma_f/\sigma_p = 1$. The purity at this point is not close to unity.\label{fig:purity_optimisation}}
\end{figure}

\begin{figure}[httb]
	\centering
	\includegraphics[width=0.8\linewidth]{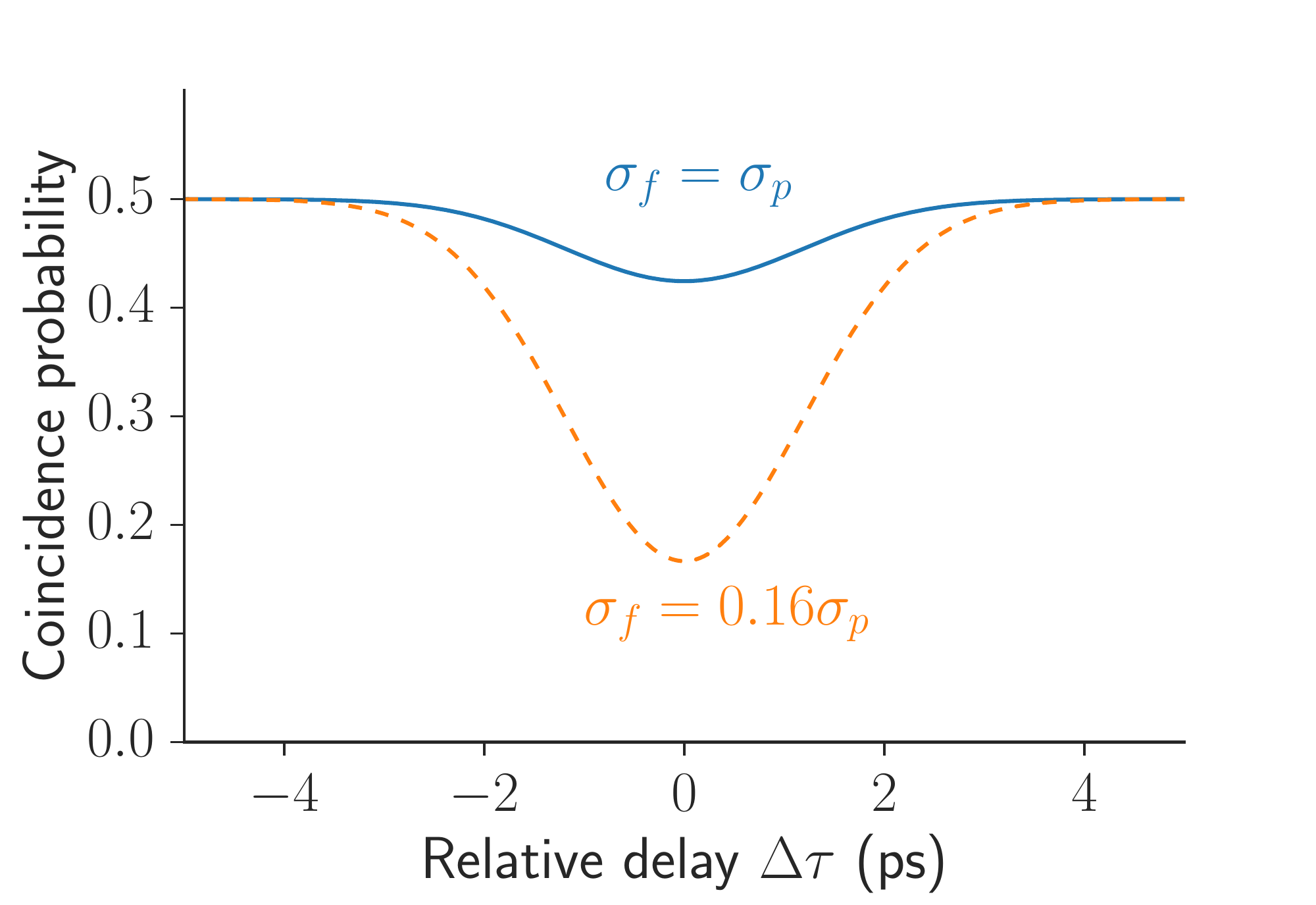}
	\caption{The coincidence probability for two identical heralded single photon sources characterised by the JSA as shown in Fig.~\ref{fig:fitted_jsa}, showing the dip characteristic of HOM interference. The solid blue line represents a filter width of $\sigma_f =\sigma_p$, leading to a visibility of only $11\%$. The orange dashed line ($\sigma_f = 0.16 \sigma_p$) yields a visibility of $50\%$.\label{fig:hom_shape}}
\end{figure}

Given a heralded photon source, one application of these results is the selection of an appropriate filter to enable highly pure heralded single photon generation at a reasonable rate. For example, we consider the source from \cite{Branczyk2010}, a periodically poled KTP waveguide. The waveguide is pumped at $\omega_p = 2\pi c/\left(\SI{400}{\nano\meter}\right)$, with a pulse duration of $\tau = \SI{0.2}{\pico\second}$. The phasematching bandwidth is 
\begin{align}
\frac{\Delta\omega_{pm}}{2\pi} =& \frac{s \pi^{-1} L^{-1}}{\sqrt{\left[k_\beta'\left(\omega_p\right)-k_\gamma'\left(\frac{\omega_p}{2}\right)\right]^2+\left[k_\beta'\left(\omega_p\right)-k_\gamma'\left(\frac{\omega_p}{2}\right)\right]^2}}\nonumber\\
=& \SI{0.13}{\tera\hertz},
\end{align}
where $s \approx 1.392$ is the first root of ${\text{sinc}^2(x) = 1/2}$, and ${k_{\beta,\gamma}(\omega)}$ are dispersion relations. Here ${k_{\beta,\gamma}'(\omega_i) = \frac{\Diff{k_{\beta,\gamma}}(\omega)}{\Diff{\omega}}\vert_{\omega=\omega_i}}$. The phasematching angle is
\begin{align}
\theta_{pm} =& \arctan\left[\frac{k_\beta'(\omega_p) - k_\gamma'(\frac{\omega_p}{2})}{k_\beta'(\omega_p) - k_\gamma'(\frac{\omega_p}{2})}\right]\nonumber\\
=& \SI{0.97}{\radian}
\end{align}
We now relate these physical quantities to the parameters of the double-Gaussian. The connections are ${\left(\sigma_1,\sigma_2,\theta_1,\theta_2\right) = \left(\Delta\omega_p/\sqrt{\ln2},\Delta\omega_{pm}/\sqrt{\ln2},\theta_p,\theta_{pm}\right)}$, where $\theta_p = \pi/4$ is the angle of the pump in the spectral domain as determined by energy conservation, and we note that $\Delta\omega_p = 1/\tau$ is the pump bandwidth (related to this we define ${\sigma_p \equiv \Delta\omega_p/\sqrt{\ln2}}$, and ${\sigma_{pm} \equiv \Delta\omega_{pm}/\sqrt{\ln2}}$). The double-Gaussian JSA with these parameters can be seen in Fig.~\ref{fig:fitted_jsa}.
Without any filtering, the purity of a heralded single photon from this source is approximately $5\%$. Intuition may suggest that the appropriate filter is one which matches the pump bandwidth $\Delta\omega_p$. However, as shown in Fig.~\ref{fig:purity_optimisation}, while there is a significant increase in the purity from $\sigma_f = 5 \sigma_p$ to $\sigma_f = \sigma_p$, the purity doesn't approach $90\%$ until $\sigma_f \lesssim 0.1 \sigma_p$. Indeed, for this source we would require $\sigma_f \lesssim 0.16 \sigma_p$ to have a HOM visibility higher than $50\%$ (see Fig.~\ref{fig:hom_shape}). Noting that $\sigma_{pm} < \sigma_p$, a smarter choice might be ${\sigma_f = \sigma_{pm} = 0.12 \sigma_p}$. Then $\mathcal{P} = 78\%$, and $V = 64\%$. Whilst this still does not reach $90\%$ purity, it is an improvement over matching the pump bandwidth in this case. 

Additionally we may place filters on both the heralding and heralded arms, with the expectation that this should improve the purity~\cite{Meyer-Scott2017a}. This is less relevant to heralded single photons as we do not wish to introduce uncertainty as to whether our heralded photon passes through its respective filter. Regardless, our method is easily extended to adding an additional filter of transmission $|t_X(\omega)|^2$ to the heralded arm. From Eq.~(\ref{eq:heralded_single_photon_state}), we project into the subspace where the idler photon has been transmitted through its own filter. In this case, the purity is given by
\begin{multline}\label{eq:two_filt_purity}
\mathcal{P}_2 = \frac{1}{\mathcal{S}_2^2} \int \Diff{\omega}\,\Diff{\omega'}\,\Diff{\omega''}\,\Diff{\omega'''} \left|t_{\widetilde{X}}(\omega')\right|^2\left|t_{\widetilde{X}}(\omega''')\right|^2\\ \times\left|t_X(\omega)\right|^2\left|t_X(\omega'')\right|^2
\Phi(\omega,\omega') \Phi^*(\omega'',\omega')\\
\times\Phi^*(\omega,\omega''')\Phi(\omega'',\omega'''),
\end{multline}
and
\begin{equation}\label{eq:two_filt_prob}
\mathcal{S}_2 = \int \Diff{\omega}\,\Diff{\omega'} \left|t_{\widetilde{X}}(\omega)\right|^2 \left|t_X(\omega')\right|^2 \left|\Phi(\omega,\omega')\right|^2
\end{equation}
Here the subscript $2$ denotes the presence of two filters, and $\mathcal{S}_2$ is the probability of both the herald and heralded photons passing through their respective filters. The Schmidt mode versions of these expressions can be found in the appendix. As shown by the dashed lines in Fig. \ref{fig:purity_optimisation}, even with two identical filters matching the pump bandwidth, the purity does not increase to $90\%$, although there is an improvement at the cost of heralding success probability.


\subsection{Schmidt mode filtering}
\noindent A natural question to ask is whether or not there are any schemes for increasing the purity that do not have such a cost to the probability of successful heralding. If we examine Eq.~(\ref{eq:measurement_operator}), we might imagine that we could have slightly more freedom (in particular phase-sensitivity), allowing us instead to write
\begin{equation}\label{eq:phase_freedom_measurement_operator}
\hat{F}'_{\widetilde{X}} = \int \Diff{\omega}\,\Diff{\omega'} t_{\widetilde{X}}(\omega) \aopd{\widetilde{X}}(\omega)\kbvac t_{\widetilde{X}}^*(\omega')\aop{\widetilde{X}}.
\end{equation}
Noting that the JSA can always be Schmidt decomposed, we can see that this would allow us to choose the filter to match a particular Schmidt function, $t(\omega)=\Theta_a(\omega)$. In this case, the heralded single photon purity is always $\mathcal{P}=1$. However to do this, we must imagine a ``phase-sensitive filter'', a device that can distinguish between $\Theta_\mu(\omega)$ and $\Theta^*_\mu(\omega)$, for any $\mu$. One possibility is a device which discretizes the input photon spectrum into $N$ frequency bins, which are passed into an $N\times N$ array of frequency-space waveguide Mach-Zehnder interferometers. The beamsplitter operation in frequency-space can be performed via Bragg-scattering four wave mixing \cite{Gnauck2006,Bell2016}. This allows for the spatial separation of each Schmidt mode, and so we can imagine if we wished to interfere this photon with another photon from a similar source, we simply need to match the spatial modes at the output of two devices in order to interfere the photons in the same Schmidt mode. Such a device could in principle have arbitrarily low loss, and thus arbitrarily high heralding success probability $\mathcal{S}$. If we only wish to select a single Schmidt mode, with careful design we can use a quantum pulse gate \cite{Brecht2014}. Both of these approaches are a way to improve the heralded single photon purity, without the large amounts of loss associated with a narrowband filter.

\section{Conclusions}
\noindent We have demonstrated two methods for characterizing the space between two well-known classes of JSAs, namely: JSAs with an aspect ratio of one (or otherwise oriented horizontally or vertically) that naturally produce heralded single photons with unit purity, and JSAs far from these conditions that require narrowband filtering to yield high purities, at the cost of the heralding success probability (and consequently the rate). Most importantly, we have found closed form expressions for the heralded single photon purity, heralding success probability and HOM interference dip shape for a broad class of JSAs. In addition to this, we have examined the underlying mode structure in filtered heralded single photon generation in order to fully understand the physics involved. We have also demonstrated that the width of the HOM dip has no relation to the width of the herald filter, which our Schmidt mode picture shows holds true for any biphoton state. Furthermore, we show that when selecting filters, matching the filter bandwidth to the pump bandwidth is not always the best choice. Finally, we have suggested other methods for improving the heralded single photon purity which do not rely on lossy filtering, such as the quantum pulse gate. This work helps one to choose an appropriate filter for a given heralded source.

\section*{Appendix}
\noindent The orthonormality and completeness conditions of the Schmidt basis are given by,
\begin{align}
\sum_{\mu} \Gamma^*_{\mu}(\omega) \Gamma_{\mu}(\omega') &= \delta(\omega-\omega'),\\
\int \Diff{\omega} \Gamma^*_{\mu}(\omega) \Gamma_{\nu}(\omega) = \delta_{\mu,\nu}.
\end{align}

The Schmidt mode picture also captures the case of two filters by direct substitution into Eq. \ref{eq:two_filt_purity} and \ref{eq:two_filt_prob}, arriving at the two-filter expressions for the purity and probability of both photons passing through their respective filters,
\begin{align}
\mathcal{P}_2 &= \frac{1}{\mathcal{S}_2^2} \sum_{\mu\nu\mu'\nu'} \sqrt{p_\mu p_\nu p_{\mu'} p_{\nu'}} Q_{\mu\nu} Q_{\mu'\nu'} Q'_{\mu\nu'} Q'_{\mu'\nu},\\
\mathcal{S}_2 &= \sum_{\mu\nu} \sqrt{p_\mu p_\nu} Q_{\mu\nu} Q'_{\mu\nu},
\end{align}
where
\begin{equation}
Q'_{\mu\nu} = \int \Diff{\omega} \left|t_X(\omega)\right|^2 \Gamma_{\mu}(\omega)\Gamma^*_{\nu}(\omega).
\end{equation}

The Schmidt functions can be found by taking advantage of the spectral theorem, plus the complete and orthonormal nature of the Schmidt functions. One construction of this is as follows: by considering the reduced density matrix elements, $\rho_1(\omega,\omega'') = \int \Diff{\omega'} \Phi(\omega,\omega') \Phi^*(\omega'',\omega')$ (and similarly for $\rho_2(\omega',\omega''')$), and inserting the definition of the Schmidt decomposition described by Eq.~(\ref{eq:schmidt_decomp}), one can show
\begin{equation}
\int \Diff{\omega''} \rho_1(\omega,\omega'') \Gamma_{\mu}(\omega'') = p_{\mu} \Gamma_{\mu}(\omega),
\end{equation}
with a similar equation for $\rho_2(\omega',\omega''')$. The eigenfunctions associated with this problem, if found, are the Schmidt functions themselves. For double-Gaussian JSAs, these are
\begin{equation}
\Gamma_{\mu}(\omega) = (-i)^\mu \sqrt{\frac{1}{2^\mu \mu! \Omega_1 \pi^{1/2}}} \exp\left(-\frac{\omega^2}{2\Omega_1^2}\right) H_\mu\left(\frac{\omega}{\Omega_1}\right),
\end{equation}
where $H_\mu(\omega)$ are the Hermite polynomials, and $\Omega_1$ is a scale constant associated with $\Gamma_{\mu}(\omega)$. A similar equation exists for $\Theta_{\mu}(\omega)$, with associated scale constant $\Omega_2$. These constants are
\begin{align}
\Omega_1 &= \sqrt{\frac{\sigma_1 \sigma_2}{\left|\sin(\theta_1-\theta_2)\right|}} \left(\frac{\sigma_1^2 \cos^2\theta_2 + \sigma_2^2 \cos^2\theta_1}{\sigma_1^2 \sin^2\theta_2 + \sigma_2^2 \sin^2\theta_1}\right)^{1/4},\\
\Omega_2 &= \sqrt{\frac{\sigma_1 \sigma_2}{\left|\sin(\theta_1-\theta_2)\right|}} \left(\frac{\sigma_1^2 \sin^2\theta_2 + \sigma_2^2 \sin^2\theta_1}{\sigma_1^2 \cos^2\theta_2 + \sigma_2^2 \cos^2\theta_1}\right)^{1/4}.
\end{align}
The eigenvalues $p_\mu$ follow a thermal distribution ${p_\mu = \bar{\mu}^\mu /(\bar{\mu}+1)^{\mu+1}}$, and satisfy ${K = \left(\sum_{\mu} p_\mu^2\right)^{-1} = 2\bar{\mu}+1}$, where we associate $\bar{\mu}$ with the mean photon occupation number per mode. They can be expressed with respect to the Schmidt number,
\begin{equation}
p_\mu = 2\frac{\left(K-1\right)^{\mu}}{\,\left(K+1\right)^{\mu+1}},
\end{equation}
which is itself expressed in terms of the parameters of the double-Gaussian as
\begin{equation}
K = \sqrt{\frac{\left(\sigma_1^2 \sin^2\theta_2 + \sigma_2^2 \sin^2\theta_1\right)\left(\sigma_1^2 \cos^2\theta_2 + \sigma_2^2 \cos^2\theta_1\right)}{\sigma_1^2 \sigma_2^2 \sin^2(\theta_1-\theta_2)}}.
\end{equation}

\bibliographystyle{apsrev-nourl}
\bibliography{clean}

\end{document}